%Paper: hep-th/9303146
%From: "Mark Wexler" <wexler@puhep1.Princeton.EDU>
%Date: Thu, 25 Mar 93 20:46:10 EST

% Use plain TeX + harvmac
% The figures are attached at the end of the text as a self-unpacking
% shell script.  Put them into a separate file and follow the
% directions.
\input harvmac.tex
%
% various definitions
%
\def\a{\alpha}
\def\b{\beta}
\def\d{\delta}
\def\k{\kappa}
\def\L{\Lambda}
\def\z{\zeta}
\def\Tr{{\rm Tr} \hskip2pt}
\def\myappendix{\global\meqno=1\global\subsecno=0\xdef\secsym{\hbox{}}
\bigbreak\bigskip\noindent{\bf Appendix}\message{(Appendix)}
\writetoca{Appendix}\par\nobreak\medskip\nobreak}
%
% list of references
%
\lref\MatMod{
F. David, {\it Nucl. Phys.} {\bf B311} (1985) 45;
V. Kazakov, {\it Phys. Lett.} {\bf 150B} (1985) 28;
J. Ambj\o rn, B. Durhuus and J. Fr\"{o}hlich, {\it Nucl. Phys.}
   {\bf B257} (1985) 433;
D. J. Gross and A. A. Migdal, Phys. Rev. Lett. {\bf 64} (1990) 717;
E. Br\'ezin and V. Kazakov, Phys. Lett. {\bf 236B} (1990) 144;
M. Douglas and S. Shenker, Nucl. Phys. {\bf B335} (1990);
D. Gross and N. Miljkovic, Phys. Lett. {\bf 238B} (1990) 217;
P. Ginsparg and J. Zinn-Justin, Phys. Lett. {\bf 240B} (1990) 333.}
\lref\Mike{M. Newman, private communication}
\lref\BoulKaz{D. V. Boulatov and V. A. Kazakov, {\it Phys. Lett.}
{\bf 186B} (1987) 379.}
\lref\BrezHik{E. Br\'ezin and S. Hikami, {\it Phys. Lett.} {\bf 283B} (1992)
203; Phys. Lett. {\bf 295B} (1992) 209.}
\lref\BIPZ{E. Br\'ezin, C. Itzykson, G. Parisi, and J.-B. Zuber,
{\it Commun. Math. Phys.} {\bf 59} (1978) 35.}
\lref\David{F. David, {\it Nucl. Phys.} {\bf B257} [FS14] (1985) 543.}
\lref\LargeD{J. Ambj\o rn, B. Durhuus, J. Fr\"{o}hlich and
P. Orland, {\it Nucl. Phys.} {\bf B270} [FS16] (1986) 457;
D.V. Boulatov, V.A. Kazakov, I.K. Kostov and A.A. Migdal,
{\it Nucl. Phys.} {\bf B275} [FS17] (1986) 641.}
\lref\Altern{E. Br\'ezin and J. Zinn-Justin, {\it Phys. Lett.}
  {\bf 288B} (1992) 54;
L. Alvarez-Gaum\'e, J.L.F. Barb\'on and C. Crnkovic, ``A proposal for
  strings at $D>1$,'' CERN preprint CERN-TH-6600-92, July 1992
  (hep-th/9208026);
M. Staudacher, ``Combinatorial solution of the two matrix model,''
  Rutgers preprint RU-92-64, January 1993 (hep-th/9301038)}
\lref\Mehta{M.L. Mehta, {\it Commun. Math. Phys.} {\bf 79}
  (1981) 327}
\lref\ItzZub{C. Itzykson and J.-B. Zuber, {\it J. Math. Phys}
  {\bf 21} (1980) 411}
\lref\IndQCD{V. Kazakov and A.A. Migdal, ``Induced QCD at large N,''
  Princeton preprint PUPT-1322, June 1992 (hep-th/9206015)}
\lref\Kostov{I. Kostov, in {\it Non-perturbative
  aspects of the standard model}: proceedings of the XIXth
  International Seminar on Theoretical Physics, Jaca Huesca, Spain,
  6-11 June 1988, eds.~J.~Abad, M.~Belen Gavela, A.~Gonzalez-Arrago
  (North-Holland, Amsterdam, 1989) 295}
\lref\GN{D.J. Gross and M.J.Newman, {\it Phys. Lett.}
  {\bf 266B} (1991) 291}
\Title{\vbox{\baselineskip12pt\hbox{PUPT-1384}
}}
{\vbox{\centerline{Low Temperature Expansion of Matrix Models}}}
\centerline{{Mark Wexler}\footnote{$^\dagger$}
{E-mail address: \it{wexler@puhep1.princeton.edu}}}
        \vskip2pt\centerline{\it{Department of Physics}}
        \vskip1pt\centerline{\it{Princeton University}}
        \vskip1pt\centerline{\it{Princeton, NJ 08544 USA}}
\rm
\vskip .5in
\noindent
We show how to expand the free energy of a matrix model
coupled to arbitrary matter in powers of the matter coupling
constant.  Concentrating on $\nu$ uncoupled Ising models---which
have central charge $\nu/2$---we work out the expansion to sixth
order for $\nu$ = 1, 2, and 3.  Analyzing the series by the ratio
method, we exhibit the spin-ordering phase transition.  We discuss
the limit $\nu \rightarrow \infty$, which is especially clear in
the low temperature expansion; we prove that in this limit the
dependence of the model on $\nu$ becomes trivial.

\Date{March 1993}

\newsec{Introduction}

One of the unsolved questions in matrix models \MatMod~is, how
do you introduce matter?  This is an important
question.  If the matrix model is viewed
as two-dimensional euclidean quantum gravity, the matter
is the statistical system which is coupled to the random
surface.
If the matrix model is viewed as bosonic string theory,
the matter represents the target space in which the string
propagates.  And at least one phase of nonabelian gauge
theory can be reduced to a matrix model with matter \IndQCD.

A general framework for coupling matter to matrix models
has not been found.  The only exactly solved cases are the
open chain of matrices, which gives unitary models with
central charge $c<1$, and the one-dimensional case.
The region $c>1$ remains almost completely unknown,
although it is easy to formulate matrix models with any
central charge.
Recently there have been a number of attempts to develop
alternate schemes that one hopes could deal with matter
in this region \refs{\BrezHik,\Altern}.  The method and
results of \BrezHik~are promising, but one is nervous
about modeling random surfaces out of about ten squares or
triangles, which are rather unwieldy, especially if
the surfaces are to be branched polymers.

In the study of conventional spin models, low and
high temperature expansions have proven valuable.
Maybe such expansions can prove useful for spin systems
coupled to lattices as well.
Before Mehta solved matrix chains \Mehta,
Itzykson and Zuber \ItzZub~suggested expanding the
two-matrix model in low temperature series.
This suggestion has been forgotten, but the
problem of coupling $c>1$ matter to a fluctuating surface
remains unsolved.
Here we develop a low temperature expansion for matrix
models.  In each order $n$ in the matter coupling constant,
we get arbitrarily large surfaces in which all the links
join equal spins, except for $n$ ``bridge'' links.
These bridges connect the several ``blobs,'' open surfaces
on which all spins are frozen equal; two blobs connected by
one or more bridges have unequal spins.  This is the random
surface analog of the ordinary low temperature expansion.
There are two advantages of this expansion:
one can expand a system with arbitrary central charge
to rather high (or infinite) order in the cosmological
constant, obtaining surfaces which are---hopefully---close
to the continuum; and one can use it to study the $c\to\infty$
limit.

\newsec{The expansion}

The partition function of a matrix model coupled to arbitrary
matter can be written
\eqn\ZQ{
Z_Q(g,a) = \int\prod_i {\cal D} \phi_i \hskip3pt
e^{-{\rm Tr}\left[\sum_i V(\phi_i) -
\sum_{ij} Q_{ij} \phi_i \phi_j\right]}
}
where $\phi_i$ are hermitian $N \times N$ matrices.
We consider cubic models, with $V(\phi) = \phi^2/2
+ g \phi^3/\sqrt{N}$.  The type of matter is encoded by the
number of matrices in the model and the matter
coupling matrix $Q$, which depends on $a=e^{-\beta}$,
the matter coupling constant.

Here we specialize to $\nu$ {\it uncoupled} Ising models.
They are ``uncoupled'' only in the bare action, though;
each one interacts with the fluctuating surface and they thus
interact among themselves.  At the critical point of
such a model, the central charge of the matter is $\nu/2$.
For $\nu$ Ising models, we need $2^\nu$ matrices to represent
every combination of spins at each site.  The coupling
matrix $Q$ can be considered as the connection matrix of
the target space graph;
the graph is a $\nu$-dimensional hypercube, in which the connection
strength between any two vertices is $a$ raised to the power
of the dimension of the lowest simplex which contains both
vertices.
For one or two Ising models,
for example, the matrices
can be written\footnote{$^\dagger$}{
A superscript in parentheses will refer to $\nu$.}:
\eqn\Qex{
Q^{(1)} = {1\over2}\left(\matrix{0&a\cr
				 a&0\cr}\right)
\hskip1cm
Q^{(2)} = {1\over2}\left(\matrix{0&a&a&a^2\cr
				 a&0&a^2&a\cr
				 a&a^2&0&a\cr
				 a^2&a&a&0\cr}\right)
}

In the spherical limit, the free energy of these models
is
\eqn\Fnu{
F^{(\nu)}(g,a) = \lim_{N\to\infty}
{1\over N^2} \log
{Z^{(\nu)}(g,a)
\over Z^{(\nu)}(g,0)}
}
(which we normalize for later convenience).  The
problem is to expand $F^{(\nu)}$ in powers of a.
We begin by expanding $Z^{(\nu)} = z^{(0)} +
z^{(\nu)}_1 a + z^{(\nu)}_2 a^2 + \cdots$; of course,
we will generate both disconnected and connected terms,
but the former will be canceled by the logarithm.
Consider $z^{(1)}_1$:
\eqn\zoneone{
z^{(1)}_1 = \langle {\rm Tr}\hskip2pt \phi_1 \phi_2 \rangle
= \langle \phi^{\alpha}_{\beta} \rangle
  \langle \phi^{\beta}_{\alpha} \rangle,
}
where in the last expression the averages are with
respect to a {\it single} matrix model.  By expressing the
traces in terms of components, we have reduced the
calculation to two surfaces (blobs), each of which has uniform
spins, with one link (bridge) joining the two.
Similarly, $z_2$ will
contain contributions from two blobs joined
by two bridges, or three blobs joined in
an open chain.  Obviously, we can repeat
this procedure for all multi-matrix model averages,
yielding contractions of one-matrix model
Green's functions.

The object of interest, then, is the one-matrix model average
tensor $\langle \phi^{\alpha_1}_{\beta_1} \cdots
\phi^{\alpha_n}_{\beta_n} \rangle$.  Because of the
$\phi \mapsto U^\dagger \phi U$ symmetry, it can
only depend on $\delta^{\alpha_i}_{\beta_j}$, and we
must keep separate the upper and lower indices
In general,
\eqn\firsttensor{\eqalign{
\langle \phi^{\alpha_1}_{\beta_1} \cdots
\phi^{\alpha_n}_{\beta_n} \rangle =
\sum_{\pi\in\Pi_n} \lambda_{n,\pi} \hskip2pt
T^{\alpha_1 \ldots \alpha_n}_{\beta_1 \ldots \beta_n}(\pi) \cr
T^{\alpha_1 \ldots \alpha_n}_{\beta_1 \ldots \beta_n}(\pi) =
\delta^{\alpha_1}_{\beta_{\pi_1}} \kern-2pt\cdots
\delta^{\alpha_n}_{\beta_{\pi_n}} \cr
}}
where $\Pi_n$ is the set of permutations of $n$
objects.
Because the matrix components
commute, many of the coefficients in
\firsttensor~are equal.  We can characterize this by a mapping
from permutations to the partitions of the integer $n$,
$f: \Pi_n \to P_n$,
where $f(\pi)$ is the set of the lengths of
the cycles of permutation $\pi$; if $f(\pi) = f(\pi')$,
then $\lambda_\pi = \lambda_{\pi'}$.  Therefore we define
\eqn\secondtensor{\eqalign{
\langle \phi^{\alpha_1}_{\beta_1} \cdots
\phi^{\alpha_n}_{\beta_n} \rangle =
\sum_{p\in P_n} \kappa_{n,p}
\left[(T^{\alpha_1 \ldots \alpha_n}_{\beta_1 \ldots \beta_n}(\pi_1) +
T^{\alpha_1 \ldots \alpha_n}_{\beta_1 \ldots \beta_n}(\pi_2)
+ \cdots \right] \cr
f(\pi_1) = f(\pi_2) = \cdots = p
}}
For $n=3$, for example, we have
\eqn\nthree{\eqalign{
\langle \phi^{\alpha_1}_{\beta_1} \phi^{\alpha_2}_{\beta_2}
\phi^{\alpha_3}_{\beta_3} \rangle = &
\k_{3,1} \d^{\a_1}_{\b_1} \d^{\a_2}_{\b_2} \d^{\a_3}_{\b_3}
+ \k_{3,2} \left(\d^{\a_1}_{\b_1} \d^{\a_2}_{\b_3} \d^{\a_3}_{\b_2}
+ \d^{\a_1}_{\b_3} \d^{\a_2}_{\b_2} \d^{\a_3}_{\b_1}
+ \d^{\a_1}_{\b_2} \d^{\a_2}_{\b_1} \d^{\a_3}_{\b_3} \right)\cr
& + \k_{3,3} \left(\d^{\a_1}_{\b_3} \d^{\a_2}_{\b_1} \d^{\a_3}_{\b_2} +
\d^{\a_1}_{\b_2} \d^{\a_2}_{\b_3} \d^{\a_3}_{\b_1} \right)\cr
}}
(for a given $n$, the $\k$'s will be numbered in lexicographic
order of the partitions).

The task now is to calculate the coefficients $\k_{n,p}(g)$.
(The reader may wonder why we  do not simply use one-matrix
model connected Green's functions \BIPZ~to represent the blobs.
We wish that this were possible, but---despite one's
intuition---blobs that make up a planar surface may themselves
be counted by nonplanar Green's functions.)
The most straightforward method is to contract \secondtensor~with
the various tensors
$T^{\alpha_1 \ldots \alpha_n}_{\beta_1 \ldots \beta_n}(\pi)$,
obtaining a closed set of linear equations for $\k_{n,p}$,
where the inhomogeneous terms are one-matrix averages of
products of traces.  This becomes quite cumbersome, though,
when $n$ gets large.
Another method is to contract \secondtensor~with
$\L^{\b_1}_{\a_1}\cdots\L^{\b_n}_{\a_n}$, where $\L$ is
some $N\times N$ hermitian tensor; this gives
\eqn\contracted{
\langle({\rm Tr} \hskip2pt \L\phi)^n\rangle =
\sum_{p\in P_n} \mu_{n,p} \k_{n,p} \Tr\L^{p_1}
\Tr\L^{p_2} \cdots
}
where $p_1, p_2, \ldots$ are elements of the partition $p$,
and $\mu_{n,p}$ is the number of different permutations
$\pi\in\Pi_n$ such that $f(\pi) = p$.

The averages on the left hand side of \contracted~can
be calculated by expanding the external field
integral
\eqn\extZ{
{\cal Z}(g,\Lambda) =
\int{\cal D}\phi \hskip3pt
e^{-{\rm Tr}[V(\phi) - \Lambda\phi]}
}
in powers of $\L$.  Fortunately, this integral has been
computed in the spherical limit
by Kazakov and Kostov \Kostov~and
by Gross and Newman \GN~using
loop equations.\footnote{$^\dagger$}
{This integral seems to be much
harder for a quartic than for a cubic potential.  That is
the reason why we use a cubic potential in this work.}
Using Gross and Newman's notation,
\eqn\KKGN{\eqalign{
{\cal F}(g,\Lambda) = \lim_{N\to\infty}
{1\over N^2} \log {\cal Z}(g,\Lambda)
= & -{1\over 2N^2}\sum_{a,b}\log(\mu_a+\mu_b)
-{1\over 6g}(\sigma_{-2}-x)
+ {2\over\sqrt{27g}} \sigma_{-3}\cr
& + \sigma_1 \sigma_{-1}
+ \sqrt{g\over 48} \sigma_1^3
- {1\over 108 g^2} - {1\over4} \log 3g \cr
}}
where $\mu_a =
\sqrt{\lambda_a + x}$, $\sigma_k = {1\over N}
\sum_a 1/(\lambda_a + x)^{k/2}$, $\lambda_a$
are the eigenvalues of $\Lambda$, and $x$ satisfies
the equation
\eqn\eqx{
x = 1/(12 g) - \sqrt{3g} \sigma_1(x).
}

We first expand $x = x_0 + x_1 \Tr\L +
x_{21} (\Tr\L)^2 + x_{22} \Tr\L^2 + \cdots$;
the correct root of \eqx~has $x_0 = 1/(12 g) - 6 g + \cdots$.
Plugging $x$ into \KKGN~and comparing with
\contracted, we can calculate the required coefficients.
There is only one subtlety: we do not want Green's
functions where two external legs are connected directly to
each other, as this would ``short circuit'' that surface;
{\it i.e.,} the number of dissident links would be one
less than required.  Since the Green's functions generated
by ${\cal F}$ are connected, the problem only occurs in
second order; therefore we must subtract the constant
term from that Green's function before we exponentiate
${\cal F}$ to obtain ${\cal Z}$.  Proceeding confidently,
we can now read off the coefficients.
The first few are:
\eqn\Kexamples{\eqalign{
\k_{1,1}&= \left(-{1\over 6g} + {1\over 4x_0}
+ \sqrt{x_0\over 3g} \right) \sqrt{N}\cr
\k_{2,1}&= {1\over 16 x_0^2 N} + \left(
-{1\over 6g} + {1\over 4x_0} + \sqrt{x_0\over 3g}
\right)^2 N \cr
\k_{2,2}&= {1\over 2\sqrt{3gx_0}} - {1\over 16x_0^2} - 1 \cr
}}
and $x_0$ satisfies $x_0 = 1/(12g) - \sqrt{3g/x}$.

Equipped with the coefficients $\k$, we can directly
evaluate the free energies \Fnu~for any value of $\nu$.
The $a^0$ term is of course just $2^\nu$ times the free
energy of the one-matrix model \BIPZ.
In the Appendix, we give $F^{(\nu)}$ through order $a^3$
where the coefficients are given for arbitrary $\nu$ as
exact functions of $\nu$ and $g$; in other words, we have
included arbitrarily large surfaces.
Beyond that, we have calculated $F^{(1)}$,
$F^{(2)}$, and $F^{(3)}$ through order $a^6$, where the
series coefficients have been expanded to order $g^{32}$,
meaning that we have included surfaces made of up to 32
triangles.  We will present a brief analysis of the series
in the next section, but first we will give a summary
of the difficulties encountered in the calculation, and
how we can check it.

When one expands the interaction term in $Z^{(\nu)}$ in
powers of $a$, in each order one gets large numbers of
graphs in target space, which is a $\nu$-dimensional
hypercube for $\nu$ Ising models.  The number of graphs
increases rapidly with both $\nu$ and the order in $a$.
Many of the graphs are isomorphic, and so give the same
results.  The choice is between generating many labeled
graphs, or many fewer unlabeled graphs.  In the latter
case (which is what we did up to order 3), however,
one has to solve the difficult combinatorical problem
of how many ways there are to embed each unlabeled graph
in the hypercubic target space.
The other major computational difficulty is in expanding
the external field integral \extZ; one could, perhaps,
speed things up by dropping the quadratic term from
$V(\phi)$ and then shifting $\phi$ to induce it~\Mike.

We can check $F^{(1)}$, of course, against the exact result
\BoulKaz.  For more than one Ising model, we can regroup
the series into powers of $g$, and compare with small-surface
expansions similar to the ones developed by Br\'ezin and
Hikami \BrezHik.  This is a series to much higher order
in $a$ but much lower order in $g$ than what we have.
One may, of course, also use the low temperature expansion
to check the small-surface expansion.

\newsec{Series Analysis}

To exhibit critical behavior, we perform a series analysis
modeled after those in \BrezHik.  Having expanded to sixth
order, we do not expect to calculate the critical exponents,
but we will learn something about the critical
behavior.  We first regroup the
series into powers of $g$.  The coefficient $A_n$ of $g^n$ is then
the sum of the matter partition functions on all the $n$-th
order graphs, {\it i.e.,} surfaces of $n$ triangles; we know
it only to order $a^6$, though.  If we had calculated it
to all orders in $a$, its asymptotic behavior at large $n$
would be
\eqn\asymp{A_n \approx g_c^{-n} n^\z}
where the exponent $\z = \gamma_{\rm str} - 3$.
{}From the exact solution for $\nu=1$ one finds \BoulKaz~that
$\z(a) = -7/2$, the pure gravity value, for all $0\le a\le 1$
except the critical point $a^{*(1)} = (2\sqrt{7}-1)/27
\doteq 0.1589$, which is where the spin ordering phase
transition takes place and where $\z = -10/3$.
This is how we will look for the phase transition.

To calculate $\z(c)$, we will use a ratio method \David.
We define $r^{[1]}_n = A_n/A_{n-2}$,
$q^{[1]}_n = n (r^{[1]}_n - r^{[2]}_n)/2$,
and for $u>1$,
\eqn\defratio{\eqalign{
r^{[u]}_n&= {n r^{[u-1]}_n - (n-2u+2) r^{[u-1]}_{n-2} \over
2p - 2} \cr
q^{[u]}_n&= {n q^{[u-1]}_n - (n-2u+2) q^{[u-1]}_{n-2} \over
2p - 2} \cr
s^{[u]}_n&= r^{[u]}_n/q^{[u]}_n \cr
}}
Naively, the asymptotic behavior as $n \to\infty$ should be
\eqn\asymprat{\eqalign{
r^{[u]}_n& \approx g_c^{-2} \left[1 + {\cal O}(n^{-u})\right] \cr
s^{[u]}_n& \approx \z \left[1 + {\cal O}(n^{-u})\right] \cr
}}
This is true, however, only if there are no confluent
singularities.  Another difficulty with the method is that
as $u$ gets large, the coefficient of the $n^{-u}$ term can
get large as well.

We start with one Ising model.
In \fig\highu{$\z(a)$ for $\nu=1$ (one Ising model), different
iterations of the ratio method: $u = 3,\ldots,7$.
The vertical line marks the known critical point.}~we
plot $s^{[u]}_{32} \approx \z(a)$
from the sixth-order series for various iterations
of the ratio method, $u = 3,\ldots,7$.  The known value of
$a^{*(1)}$ is shown as a vertical line.  At $a=0$, $\z$ is very
close to its exact value for pure gravity, $-7/2$; for $u=4$,
for example, we have $\z \doteq -3.4991$.  These graphs are not
exactly what we would have expected; the peak, for one, is much
too high.  Nonetheless, \highu~does give qualitative evidence
for the spin-ordering phase transition.  Moreover, as we increase
$u$, we obtain increasingly accurate values for $a^*$ (the closest
one being $0.1547$ at $u=6$),\footnote{$^\dagger$}
{One can also get very
good estimates for $a^*$ from the peak in the specific heat,
$C \approx {d^2 \over da^2} \log g^*$.}
as well as lower peaks.  The ratio
approximants seem to deteriorate, however, after $u=4$, developing
a second peak.  We will therefore use a compromise value, $u=4$.

\nfig\highcone{$\z(a)$ for $\nu=1$, to orders
$a^3, a^4, a^5,$ and $a^6$}
\nfig\highctwo{Same as \highcone, $\nu=2$}
\nfig\highcthree{Same as \highcone, $\nu=3$}
In \figs{\highcone{--}\highcthree}, we plot $\z^{(\nu)}(a)$
for $\nu=1, 2,$ and 3, keeping terms up to orders 3--6
in $a$.
The right-hand side of the graphs is given only
for completeness, as there is no reason to trust
the expansion when $a$ is not small; if we had expanded
to all orders in $a$, the curve would come back to
$-7/2$ at $a=1$, because at infinite temperature
each spin fluctuates independently.
Two things can be claimed with certainty: the plots
are showing evidence for the spin-ordering phase
transition which becomes stronger and more realistic
as we expand to higher orders in $a$; and the critical
point gets closer to zero as the central charge
increases.

\newsec{The $\nu\to\infty$ limit}

As can be seen for the first three orders from the expression
for the free energy in the Appendix, apart from an overall
normalization, the coefficient of $a^n$ in the free energy
is an $n$-th degree polynomial in $\nu$.  This property,
although not a priori obvious, can be shown to hold to all
orders.  The easiest way to see it is through the small-surface
expansion \BrezHik.
There, the coefficient of $g^m$ in the free energy is a
sum of the (one Ising model)
partition functions of all surfaces of area $m$,
with each partition function raised to the power $\nu$.
Each of those partition functions is a $3m/2$-degree polynomial
in $a$, so its $n$-th power cannot have any higher power
of $\nu$ than $\nu^n$.  This proves the assertion.

We can see immediately that the asymptotic dependence
of ${\cal F}^{(\nu)}$ on $\nu$
becomes trivial as $\nu$ (and therefore the central
charge) approaches $\infty$,
since $\nu$ will then simply be a multiplicative renormalization
for $a$.  The critical exponents of the system will be
independent of $\nu$, and the critical temperature will have
the asymptotic behavior
\eqn\casymp{
a^{*(\nu)} \to {a_\infty \over\nu} \quad
{\rm as}\quad \nu \to \infty.
}
This is already approximately true in our results for
up to $\nu=3$.  In \fig\asympfig{The critical point
$a^{*(\nu)}$ for $\nu=1, 2,$ and 3}~we plot
$a^{*(\nu)}$ for $\nu=1, 2,$ and 3, showing the
plausibility of the asymptotics \casymp.
It is also intriguing to note that in the third-order
expression for the free energy in the Appendix,
the graphs which appear as coefficients of $\nu^n$ in
order $n$ are all trees.  If this behavior can be
shown to hold in all orders, this will prove that
surfaces in the $c\to\infty$ limit are branched
polymers.

\newsec{Discussion}

In this letter, we have shown how to expand a matrix model
coupled to arbitrary matter in powers of the matter coupling
constant.  We have shown that already at order $a^6$ one
can qualitatively observe the spin-ordering phase transition,
with good quantitative results for the critical coupling.
The immediate extension to this work would be to expand
to higher orders (in $a$ and $g$), and to refine the
series analysis.  The reader should know that the present
work was carried out with rather simple algorithms
programmed in the useful but very slow {\it Mathematica}
language, running on an overloaded Iris 4D/480S; altogether, the
calculations took about a day of real time to execute.
For conventional spin models, the low temperature series
are typically known to order 20 or 30, which gives accurate
results for exponents; perhaps one could push the
present expansion as far?

Another possibility is to expand other thermodynamic
quantities than the free energy.
The magnetic susceptibility series,
for instance, is known to converge rapidly.  It is not hard
to add a magnetic field to a matrix model \BoulKaz.
One could, of course, also experiment
with different types of matter such as Potts models.

Finally, the limit $\nu\to\infty$ is worth studying.
This limit becomes quite clear in the context of
the low temperature expansion: one must calculate
the $\nu^n$ coefficient of the $a^n$ term.
Many of the calculations leading to the low temperature
series drastically simplify in this limit.  The combinatorics
on a hypercube, for example, is much simpler when its
dimension goes to infinity.
There exist predictions for behavior of a random surface
embedded in $D\to\infty$ dimensions \LargeD~which can
be tested.  In this way,
perhaps we could learn whether the central charge $c\to\infty$
limit is universal.

\myappendix

Here we give the free energy for arbitrary $\nu$ to first
three orders.\footnote{$^\dagger$}{As in eq. \Fnu, the expression
here does not include the homogeneous (pure gravity) term.}
$\k_{n,p,\ell}$ is the coefficient of the
$N^\ell$ term in $\k_{n,p}$.
\eqn\appF{\eqalign{
{F\over 2^\nu} = & {1\over2} \,\k_{1,1,{1 \over 2}}^2 \, \nu \, a\,\,+\cr
& {1\over4} \Bigr[ \left(\k_{1,1,{1 \over 2}}^2 +
  2\k_{1,1,{1 \over 2}}^2 \k_{2,1,-1}
  + 2\k_{1,1,{1 \over 2}}^2 \k_{2,2,0}\right) \nu^2\cr
  &\,\quad\left(\k_{2,2,0}^2 -
  \k_{1,1,{1 \over 2}}^2\right)\nu\Bigr] a^2\,\,+ \cr
& {1\over12} \Bigl[ \k_{1,1,{1 \over 2}}^2 \Bigl(1 +
  6\left(\k_{2,1,-1}+\k_{2,2,0}\right) +
  6\left(\k_{2,1,-1}+\k_{2,2,0}\right)^2\cr
  &\qquad\quad+2\k_{1,1,{1 \over 2}}(\k_{3,1,-{5 \over 2}} +
  3\k_{3,2,-{3 \over 2}} + 2\k_{3,3,-{1 \over 2}})\Bigr)\,\nu^3\,+\cr
  &\qquad + 3\k_{1,1,{1 \over 2}} (2\k_{2,2,0}\k_{3,2,-{3 \over 2}}+
  4\k_{2,2,0}\k_{3,3,-{1 \over 2}}-\k_{1,1,{1 \over 2}} \cr
  &\qquad\quad -2\k_{1,1,{1 \over 2}}\k_{2,1,-1}-
  2\k_{1,1,{1 \over 2}}\k_{2,2,0})\,\nu^2\,+ \cr
  &\qquad 2(\k_{1,1,{1 \over 2}}^2 + \k_{3,3,-{1 \over 2}}^2)\, \nu
  \Bigr] a^3 + {\cal O}\left(c^4\right) \cr
}}
The $\k_{n,p}$ for $n\le0$ were given in the text;
the third-order coefficients are:
\eqn\Kthree{\eqalign{
\k_{3,1} = & \left({3g + 2 x_0\sqrt{3gx_0}\over 8 x_0^3
(3g - 4 x_0^3)}\right)N^{-5/2} \cr
&\qquad + 3 \k_{1,1,{1 \over 2}} \k_{2,1,-1} N^{-1/2}
+ \k_{1,1,{1 \over 2}}^3 N^{3/2} \cr
\k_{3,2} = & -{N^{-3/2} \over 16 x_0^3} +
\k_{1,1,{1 \over 2}} \k_{2,2,0} N^{1/2} \cr
\k_{3,3} = & \left({1\over32x_0^3} -
{1\over 8 x_0 \sqrt{3gx_0}}\right) N^{-1/2} \cr
}}

\listrefs
\listfigs
\end